\documentclass[useAMS,usenatbib]{mn2e}
\usepackage{graphicx}
\usepackage[compact]{titlesec}
\usepackage{amsmath}
\titlespacing{\section}{0pt}{3.5ex}{1.5ex}

\title[Polarimetric Tomography of Blazar Jets]{Polarimetric Tomography of Blazar Jets}
\author[U. Barres de Almeida et al.]{Ulisses Barres de Almeida,$^{1}$\thanks{E-mail:ulisses@cbpf.br} Fabrizio Tavecchio$^{2}$ and Nijil Mankuzhiyil$^{3}$\\
$^{1}$Centro Brasileiro de Pesquisas F\'{i}sicas, Rua Dr. Xavier Sigaud 150, Urca, Rio de Janeiro, RJ 22290-180, Brazil\\
$^{2}$Osservatorio Astronomico di Brera, Via Emilio Bianchi 46, 23807 Merate, Lecco, Italy\\
$^{3}$INFN Trieste and University of Udine, via delle Scienze 208, Udine 33100, Italy;\\ Astrophysics Science Division, BARC, Mumbai, 400085, India. 
}

\begin{document}

\date{Accepted. Received ; in original form }

\pagerange{\pageref{firstpage}--\pageref{lastpage}} \pubyear{2002}

\maketitle

\label{firstpage}

\begin{abstract}
In this paper we propose a way to use optical polarisation observations to provide independent constraints and guide to the
modelling of the spectral energy distribution (SED) of blazars, which is particularly useful when two-zone models are required
to fit the observed SED. As an example, we apply the method to the 2008 multiwavelength campaign of PKS 2155-304, for
which the required polarisation information was already available. We find this approach succesful in being able to simultaneously describe 
the SED and variability of the source, otherwise difficult to interpret. More generally, by using polarisation data to 
disentangle different active regions within the source, the method reveals otherwise unseen correlations in the 
multiwavelength behaviour which are key for the SED modelling.

\end{abstract}


\section{INTRODUCTION}

Extragalactic jets are gigantic structures up to hundreds of kpc across, produced when highly collimated plasma is ejected 
at relativistic speeds from the nucleus of active galaxies (AGN).The AGN system has a strongly non-isotropic radiative output, with a system of relativistic jets emanating at opposite diections from the central engine which produces a collimated emission pattern in the flow direction. When the observer's line-of-sight happens to be aligned with the outflow direction, the object is called a blazar~\citep{urr95}. The radiation Doppler boost resulting from this chance alignment renders blazars the most extreme of all AGN and the dominant extragalactic source class in the sky above 100 GeV~\citep{hin09}.    

Modelling of the broadband spectral energy distribution (SED) is a well established technique to study the physics of 
blazars~\citep{ghi98, tav98}. The general double-hump structure of the SED of blazars is well explained as synchrotron
and inverse-Compton radiation from a highly energetic population of electrons. Evidence from multiwavelength (MWL) 
observations, from radio to $\gamma$-rays, shows that the AGN emission is variable at all frequencies, and 
contemporaneous measurements indicate the SED is remarkably correlated~\citep{fos98}. 

Although one-zone models~\citep{mar92, sik94, blo96} have been very successful in describing the SED of blazars, recent 
MWL observations, with better temporal resolution and more complete spectral coverage, have revealed the necessity of 
adopting multi-zone models at least in some cases~\citep{ale12, abr12}. Given the degeneracy and large number of parameters
present in these inhomogeneous SED models, some independent insight into the properties of different particle 
populations that are simultaneously contributing to the observed emission would be decisive to better motivate and provide
additonal constraints to models.

The polarised emission from blazars was discovered early on in the observations of these objects, as the signature of 
synchrotron radiation from a non-thermal distribution of relativistic particles~\citep{ang80}. Since then it has
been an important technique to study the physics of blazar jets, allowing to probe the state of the magnetic field and 
particle populations at the emission sites as well as aspects of the source structure~\citep{bri86, jon88, lyu05}. 
Recently, a number of campaings have detected episodes of large and smooth rotation of the polarisation angle of blazars 
far in excess of 180 degrees, not compatible with turbulent or random behaviour~\citep{dar09, mar13}. These have been 
interpreted as the signature of a ubiquitous, large-scale magnetic field component, with specific geometry, responsible 
for the flow collimation and acceleration~\citep{mar08, abd10}. In radio, milli-arcsecond resolution polarisation 
interferometry allows for a mapping of the magnetic field state along the jet and the possibility to physically locate the
active emission regions~\citep{dar09, agu11}.

In this paper we present an original approach to the modelling of blazar SEDs which uses information provided by optical
polarimetric observations to put additional constraints and guide the model fits with independently-motivated
physical inputs. The optical polarimetric analysis at the basis of the technique~\citep{bar10} (henceforth BA10) allows
to identify the single component which contributes the most to the source activity in optical polarisation, separating it from
the rest of the ``quiescent jet.'' This is equivalent to deciding if an one- or two-zone model must be adopted to explain 
the polarisation behaviour in the optical band. In addition, the polarisation analysis provides a description of the 
characteristics of the ``active'' and ``quiescent'' components, such as their individual contributions to the total flux, 
which is in turn used to provide further constraints to the multi-zone SED model. 

The term {\it ``polarimetric tomography''} stresses that such use of the data has the capacity to 
disentangle some of the source's internal structure, even when it is impossible to resolve individual zones via direct imaging.
In this paper we are concerned with a description of the technique and its potential, as presented in Section 
2. A case study is shown in Section 3, where we apply the method to the modelling of the quiescent-state SED of 
PKS 2155-304. There we present an alternative, self-consistent two-zone solution to the SED, motivated by a previously undetected correlation in the dataset revealed by this approach.

\section[]{POLARIMETRIC TOMOGRAPHY}

Even if the inner jets of blazars cannot be resolved in optical waveband, polarisation observations are an
avenue to probe their internal structure. Turbulence is long believed to dominate the plasma flow at small 
scales~\citep{moo82}. Plasma turbulence reflects in the state of the magnetic field, resulting in a tangled 
structure which reduces the mean source polarisation to a few percent and imprints randomness to the source behaviour, 
except when some mechanism is at play that (if only temporarily) imparts order to the field, either locally (e.g., by shock 
compression;~\cite{lai80, hug89}) or globally (e.g., a toroidal or helical B-field configuration;~\cite{nak01, lov02}). 
Moreover, when internal shocks, for example, enhance the emissivity of a portion of the jet, the polarisation of this active 
zone can dominate that of the entire source, adding coherence to it. 

Since the polarisation is usually quite high in the active sites, such zones can dominate the source emission as seen in 
polarised light even when the photometric output of the region is far less than that from the rest of the 
``quiescent'' jet. When this is the case, we are in a situation where the polarisation has the potential to probe and 
disentangle the emission of an otherwise invisible, but very active sub-structure of the jet -- hence the term 
{\it ``tomography.''} 

When applied in itself, the polarisation analysis allows to draw a simple, two-zone model of the state of the source in 
optical. By analysing the variability of the Stokes parameters of the source, one can evaluate if it is
best described by the evolution of a single, dominant component, or if the superposed contribution of multiple varying polarised regions needs to be considered to explain the temporal behaviour of the data~\citep{hag08}. If the temporal changes of the Stokes Q and U parameters is dominated by the evolution of a single polarised region, then its polarisation quantities (polarisation degree and angle) can be derived (see Section 3.1 of BA10). Once this is established, one 
can use equations 1-2 in BA10 to model the total observed polarisation from the source as the result of the action of this changing component, superposed on the less active, but still polarised remainder of the jet. The jet structure and behaviour in optical is now interpreted as the 
combined emission of two components whose properties and temporal evolution are constrained by the polarisation data.  

The real potential of the analysis is nevertheless unveiled in a MWL context, as it introduces new and independetly motivated information with which to construct a model for the SED. Since one can expect the variable polarised region to likely be active at higher frequencies, characterising it at 
low energies can be key to understanding the SED behaviour. As will be shown next, with this technique we can follow 
independently the SED of the alleged active zone, identifying what is its individual contribution to the multiband source
behaviour, thus better constraining an otherwise degenerate two-zone SED model.    
 
\begin{figure}
\vspace*{-2.6truecm}
 \includegraphics[width=0.55\textwidth]{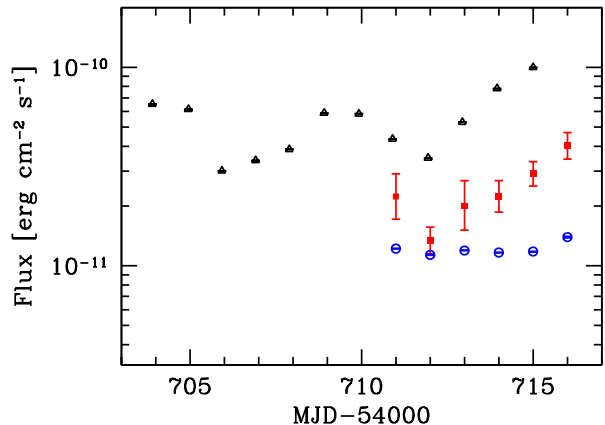}
\vspace*{-2.8 truecm}
 \caption{Black triangles show the {\it RXTE} (2-10 keV) lightcurve for PKS 2155-304 in 2008. Red squares represent the optical flux (in  $\nu F_\nu$) behaviour of the variable component as derived by BA10. Note the correlation between the source's total X-ray flux and the variability of the active optical component. The optical flux for the variable component is from a reanalysis of the data presented in BA10, following the same procedures presented in that paper but with improved numerical accuracy to the analysis. The blue open circles show the total optical flux light curve (multiplied by a factor 0.1 to fit the image scale) that does not show the same clear correlation with the X-ray data. Error bars represent the 1-sigma confidence intervals.}
\label{lc}
\end{figure}

\section{APPLICATION TO PKS 2155-304}

In September 2008, the VHE blazar PKS 2155-304 was jointly observed by H.E.S.S. and Fermi in an extensive multiwavelength 
campaign accompanied also with X-ray and optical data~\citep{aha09}. The source was found to be in a low state throughout 
the period. During the second half of the campaign, between MJD 54710-16, simultaneous optical polarimetric data was taken
 (BA10). The gamma-ray light-curves were essentially constant throughout and the optical flux showed only modest variability. The X-ray flux has in the other hand shown significant variations which could not be clearly correlated to the behaviour seen in any of the other bands, rendering the interpretation of the dataset difficult.~\cite{aha09} interpreted the source emission as a single-zone synchrotron self-Compton (SSC) model, but noted that the one-region scheme could not explain both the SED and its variability.

During the epochs of intense X-ray variability, with no signatures in gamma-ray or optical photometry, it was neverthless 
observed that the polarisation of the source was variable. A detailed analysis of the data by BA10 showed that the
polarisation behaviour could be interpreted as the interplay between two zones: a variable component and the steady 
jet. In that scenario, the photometric flux changes from the variable zone would be completely hidden by the larger flux
of the broader jet, the active region being responsible for only $\sim 10$\% of the total flux. When 
the variable zone behaviour is compared with the previously unexplained X-ray data,
a clear correlation is seen between the two (see Figure 1), suggesting that the X-ray variability could be due to 
variations of this active portion within the steady-emission jet. The previously unidentified optical properties of this active 
region, now probed thanks to the polarisation analysis, gives us the missing constraint for performing a self-consistent fit 
of the entire SED, interpreted as an independently-motivated two-zone model.

In the following section we present the SED fit performed for all nights of the campaign with simultaneous polarisation observations, putting forward a new scenario with which to interpret this data.

\begin{figure}
\vspace*{-2.0truecm}
  \includegraphics[width=0.5\textwidth]{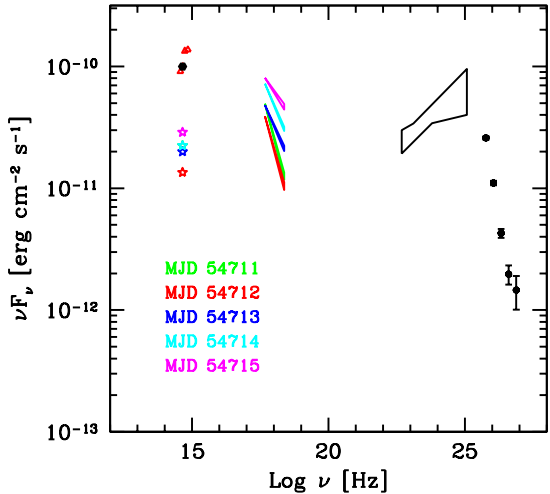}
\vspace*{-1.5 truecm}  
\caption{Contemporaneous SED of PKS 2155-304 from September 2008~\citep{aha09}. In black (filled squares) is shown the H.E.S.S. gamma-ray data. The total optical photometry is shown by the red triangle, whereas the black optical point (filled circle) is the steady component attributed to the stationary jet. Red hollow triangles are total flux photometric points for different source states. The sequence of color coded optical points (stars) represent the daily flux of the variable component which have a one-to-one correspondence with the X-rays (RXTE).}
\label{sed}
\end{figure}

\subsection{The SED modelling}

To reproduce the SED of each component we adopt a standard leptonic model~\citep{tav98}. The emission region is described 
as a sphere of radius $R$, filled with magnetic field of intensity $B$ and relativistic electrons following a smooth, 
broken power law energy distribution

\begin{equation*}
N(\gamma) \sim \gamma^{-n_1}\left(1+\frac{\gamma}{\gamma_b}\right)^{n_1-n_2}~,~~~\gamma_{\rm{min}} < \gamma < \gamma_{\rm{max}},
\end{equation*}

\noindent between Lorentz factors $\gamma_{\rm min}$ and $\gamma_{\rm max}$, with slopes $n_1$, $n_2$ below and above the break at $\gamma_{\rm b}$. Relativistic effects on the observed radiation are fully 
described by the Doppler factor $\delta$. 
Note that, from the point of view of the radiative properties, a spherical geometry is an acceptable approximation of a cylinder (expected to suitably model post-shock regions in the jet), as long as the height of the cylinder is roughly comparable to the radius. In the case in which the height reflects the cooling length of the electrons, $d\sim ct^{\prime}_{\rm cool}$, it can be checked {\it a posteriori} that the condition is fulfilled.

The optical polarisation analysis described before suggests that, at least in optical, the data set is well described by a
two-zone model, composed of a steady component, and a variable one, responsible for the variability 
seen in polarised light, but only for 10\% of the total photometric flux. The observed correlation found in Figure 1 (see also Figure 2), between the X-ray variability and the optical flux of the variable component suggests that the X-ray changes are driven exclusively by the behaviour of this portion of the jet, thus providing a tight physical constraint to the two-zone SED fit. 

In principle, these two regions could be spatially distinct or, alternatively, the system could be composed by a compact zone embedded into the larger jet, as already envisaged to reproduce the ultra-fast variability displayed by PKS 2155-304 -- e.g.,~\cite{ghi08},~\cite{gia09}. For simplicity, we do not consider this scenario here, also because it would involve the complex treatment of the radiative interplay between the two zones, difficult to include in the $\chi^2$ procedure we used (see below). Therefore we treat each component as independent and assume that within each region the inverse-Compton component derives only from the scattering of the locally produced synchrotron photons. From the point of view of the emission model this assumption is enough to fully specify the system, since for each region it allows us to derive the emitted SED separately, applying the~\cite{tav98} model. From the physical point of view, if the two regions have comparable bulk Lorentz factors (as we find to be the case) there is no relative beaming of the two radiation fields. Therefore (considering also that the two components have more or less comparable synchrotron luminosities), to ensure that the radiation field of the other component does not play an important role, it is enough that the distance separating the two regions is a few times their radii.

To reproduce the observed SED we adopted the following strategy: first of all, we fix the steady component, constrained by the steady optical flux and by the LAT and H.E.S.S. spectra. A further strong constraint to this component derives from the condition that its contribution to the X-ray emission must be very low in order to preserve the correlation shown in Figures 1 and 2. These conditions are stringent enough to provide tight constraints to the spectral shape and the model parameters for the steady component. In practice, what was done was to assume that the flux at the Fermi peak spectrum position and $\sim$ 90\% of the peak optical flux (derived from the polarimetric analysis) are coming from the steady component alone, equal at each night, allowing for 10\% variability which is permitted by the data. Then, fixing\footnote{About this choice of fixing $\gamma_{min}=1000$, we refer the reader to~\cite{sir11}, where the authors show that the electrons are pre-accelerated to $\gamma_{min} \sim m_p/m_e$.}  $\gamma_{\rm min}$=1000, a $\chi^2$ fit of the steady SED was produced following the procedure of~\cite{man12}. The values shown in the first line of Table 1 are the rounded up parameters of the fit of the steady SED fit for each day. 

Once the steady SED is fixed, we started to reproduce the variable emission, restricted by the daily optical and X-ray measurements. To fit the variable component of the SED and derive the SSC model parameters we followed the same $\chi^2$ minimization procedure of~\cite{man12}, in which the minimum of the $\chi^2$ is calculated using the {\it sum} of the variable and the quiescent component. In the fitting procedure, the last gamma-ray data point shown suffers from pile-up and thus, despite being present in Figure 3, it was not included in the fits. 

In the minimisation procedure, we assumed that the size of the emission region remained constant throughout the period, hence fixing the radius of the variable blob as $R=1.5\times 10^{16}$\,cm, which is a typical value used in the literature~\citep{tav10}. The radius of the constant component instead was produced as a result of the numerical fit, by assuming the optical and Fermi peaks as being dominated by the steady component. Due to the lack of high frequency interferometric observations during the period, it is difficult to constrain $\gamma_{\rm min}$. We have therefore opted to fix $\gamma_{\rm min}$=1000. Please, observe that, in the light of Fermi first order shock acceleration in relativistic shocks, a value of $\gamma_{min}$ much lower than this is not a good assumption, because in order for the electrons to be accelerated to very high energies they need to cross the shock several times, which in turn means they need to move faster than the shock itself, already relativistic. However, other cases have been checked (e.g., $\gamma_{\rm min}$=1, as used in~\cite{man11}), and we found that our assumption on $\gamma_{\rm min}$ does not significantly impact on the determination of the other parameters. Finally, we kept all the other 7 parameters to float in the minimization procedure, to avoid any possible bias to the final solution. The resulting fitted parameters are shown in Table 1. The $\chi^2$ minimization procedure described allowed us to rigorously calculate errors on the fit parameters; all reported (asymetric) errors are the 90\% confidence intervals for the fits, as in~\cite{man12}, and are quite small except for $\gamma_{\rm max}$ on MJD 54714, which is less constrained but still compatible with the range of this parameter for the other nights. The resulting fit parameters validate a scenario where the difference in the daily flux states result from changes in $B$, and are accompanied by some evidence of contemporaneous changes in the electron density, $K$. Along with these, changes were registered in the high-energy spectral slope $n_2$, which shows a fairly steady decrease during the campaign, and we found as well some evidence for possible changes in the break Lorentz factor $\gamma_b$, as discussed in more detail in the following section.

\begin{figure*}
 \vspace*{-0.5 truecm} 
  \includegraphics[width=0.8\textwidth]{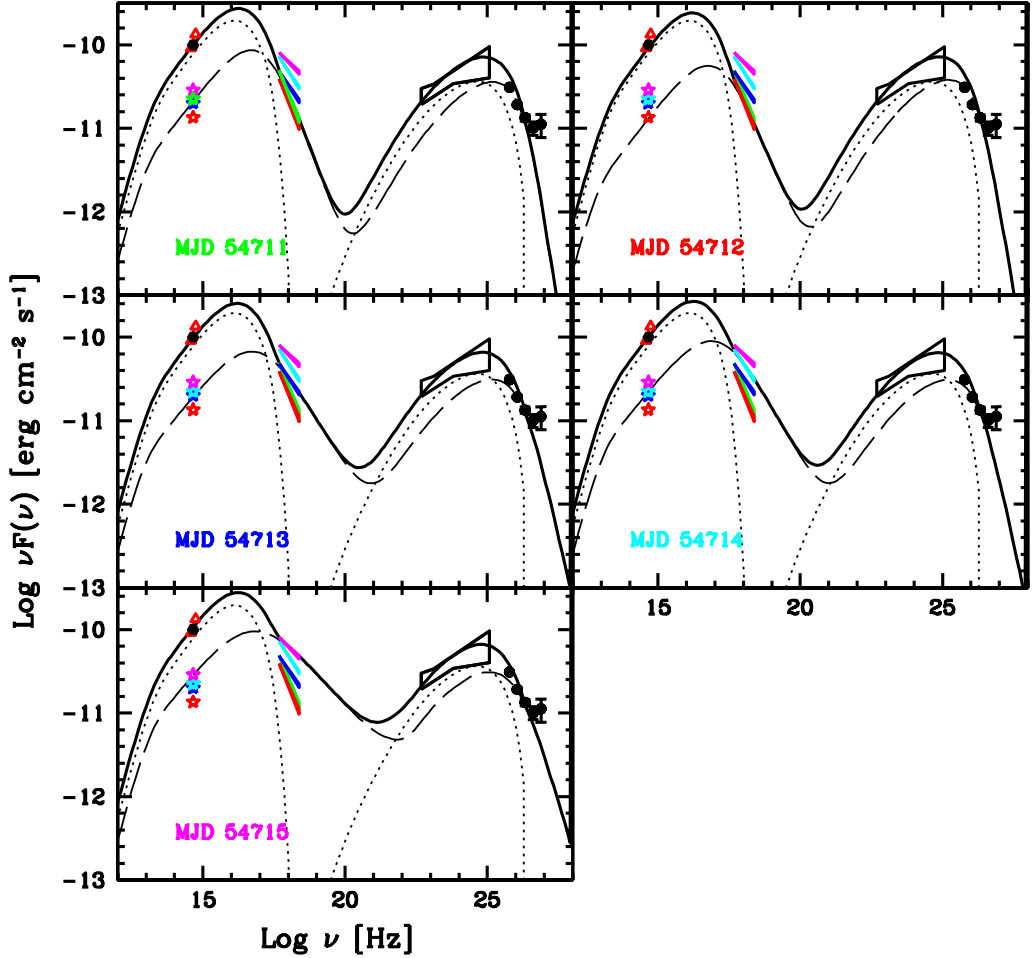}
 \vspace*{-0.5 truecm} 
  \caption{Nightly fits for the SED of PKS 2155-304. The notation for the symbols is the same as in Figure~\protect \ref{sed}. The VHE data have been EBL-de-absorbed according to~\protect \cite{dom10}. The correspondence between optical flux of the variable component and X-ray state is colour coded at each panel. The dashed line represents the SED model of the variable region, responsible for the correlated optical and X-ray changes. Dotted lines represent the SED of the steady jet, main contributor to the non-variable $\gamma$-ray flux. The total SED is shown with the thick black line as the sum of the two components. The night MJD 57111 is omitted for space constraints, but the fit is quoted in Table 1.}
\label{fits}
\end{figure*}

\subsection{Physical discussion of the results}

The introduction of optical polarimetric information into the SED analysis allowed us to derive a 
self-consistent and independently motivated scenario for the SED and its temporal evolution (see Figure~\ref{fits}), based on a new correlation observed between the X-ray data and the polarised optical flux. 
Table~\ref{param} shows the resulting parameters for the fits of the two components for each night, showing the parameters for the steady component in the first row, followed by the daily fits. 
 
\begin{figure}
\vspace*{-0.6truecm}
 \includegraphics[width=0.55\textwidth]{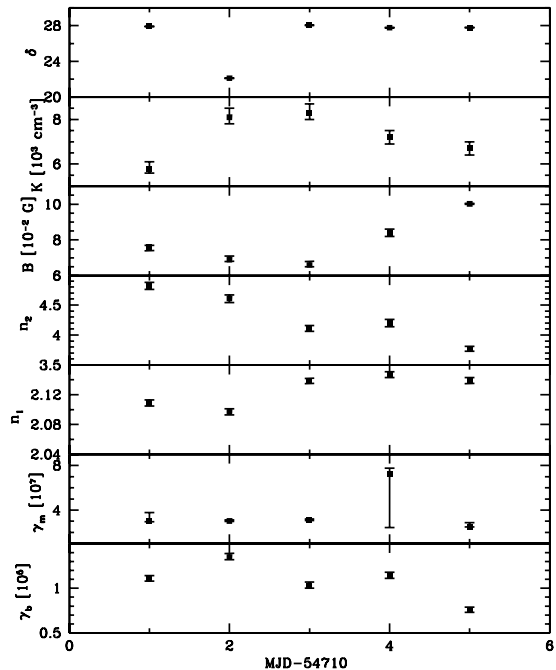}
\vspace*{-0.6 truecm}
  \caption{Plot showing the temporal evolution of the four main parameters which drive the evolution of the SED of the variable component of PKS 2155-304 in the five nights between MJD 54711-15, namely the electron density $K$, the magnetic field intensity $B$, the particle index $n_2$ and the break Lorentz factor of the electron population $\gamma_b$. The values of all parameters are shown in Table 1. Shown here are the 1-sigma errors for the parameters.}
\label{panels}
\end{figure}

Consistently with the scenario based on the polarisation analysis in BA10, in which the intensity and the degree of ordering of the magnetic field is shown by those authors to gradually increase during the campaign, we succeeded in reproducing the SEDs of the five nights for which polarisation data is available by an evolving magnetic field and particle density. As shown in Figure~\ref{panels}, while $K$ undergoes a large increase in the first night of the campaign, corresponding to an episode of particle injection at the start of the SED evolution, $B$ remains relatively unchanged at the beginning, picking up a growing towards the final dates of observations. In terms of the SED behaviour, the large jump in $K$ at the start of the campaign happens due to the increase in the optical flux observed. It is clear from Figure~\ref{panels} that $K$ is anti-correlated with $B$ throughout the campaign, and the reason for it lies in the evolution of the synchrotron and IC luminosities of the source. Despite the steady increase recorded in the synchrotron luminosity of the variable component, as constrained by the X-ray data and the polarimetric analysis, the IC peak remains essentially unchanged during the observations, as indicated by the H.E.S.S. data. The SED model achieves this balance by a gradual increase in the magnetic field, which drives up the synchrotron luminosity to its peak value at MJD 54715, but this is compensated by a correspondent decrease in the particle density in order to damp the increase in the inverse-Compton flux.

There are still other two parameters which follow a well-defined trend during the sucessive nights of the campaign. First, a gradual, monotonic hardening of the slope $n_2$ is seen, as well as a gradual decrease of $\gamma_{\rm b}$. If a hardening of $n_2$ requires that the high-energy end of the electron population be continuously replenished, the behaviour of $\gamma_b$ is consistent with a radiative equilibrium evolution for the particle population. In fact, the radiative cooling time of electrons with $\gamma \sim 10^{5}$ for $B \sim 0.1$G is just 1 day (in the observer's reference frame), and consistent with the timescales of change in the break Lorentz factor which we register.

As said before, the SED model thus built is fairly well-determined. In fact, once the steady component was fitted and fixed, constrained by the optical/X-ray flux correlation, the fit of the full SED converged to a preferred $\chi^2$-minimisation solution at each night which together form a coherent scenario for the evolution of the source behaviour. This ability to provide strong physical bounds to the two-zone model, which are usually very degenerate and where it is often difficult to find clear physical motivation for deciding among different plausible fits, is what we consider to be one of the most attractive characteristics of the technique presented here. 

We would like to advance here that, although the $\chi^2$ method was used as a means to provide an indication of how well one could constrain a solution to the two-zone model, the technique does not depend on this specific procedure to fit the SED, and is an indepent 
new approach that can give additional constraints to a two-zone SED model by using polarisation data.

\begin{table*}
\centering
\begin{tabular}{lccccccccc}
\hline
\hline

State     & $\gamma _{\rm min}$ & $\gamma _{\rm b}$  & $\gamma _{\rm max}$ & $n_1$ &$n_2$ &$B$ &$K$ &$R$ & $\delta $ \\
          & [$10^3$]          & [$ 10^5$]        &[$ 10^6$]  &  & &[$10^{-2}$G] & [$ 10^3$ cm$^{-3}]$  & $[10^{16}$ cm] &  \\
\hline
Steady    & $1.0$ & $0.4$         & $0.1$ & $2.0$ & $4.5$        & $10.0$         & $1.0$ & $2.0$ & $35.0$\\
\\
MJD 54711 & $1.0$ & $1.11^{0.03}_{0.03}$ & $30.1^{7.7}_{0.5}$ & $2.109^{0.004}_{0.004}$ & $4.82^{0.06}_{0.06}$ & $7.6^{0.1}_{0.2}$  & $5.8^{0.3}_{0.2}$ & $1.5$ & $27.90^{0.03}_{0.03}$      \\
\\
MJD 54712 & $1.0$ & $1.35^{0.04}_{0.03}$ & $30.3^{1.0}_{0.4}$ & $2.097^{0.004}_{0.004}$ & $4.61^{0.06}_{0.07}$ & $6.9^{0.2}_{0.1}$  & $8.1^{0.4}_{0.3}$ & $1.5$ & $22.12^{0.03}_{0.05}$      \\
\\
MJD 54713 & $1.0$ & $1.03^{0.04}_{0.03}$  & $31.0^{1.0}_{0.4}$ & $2.139^{0.003}_{0.004}$ & $4.11^{0.05}_{0.05}$ & $6.6^{0.2}_{0.1}$  & $8.3^{0.4}_{0.3}$ & $1.5$ & $28.05^{0.03}_{0.03}$      \\
\\
MJD 54714 & $1.0$ & $1.14^{0.04}_{0.03}$  & $72.2^{5.2}_{48.0}$ & $2.147^{0.004}_{0.004}$ & $4.20^{0.06}_{0.06}$ & $8.4^{0.2}_{0.2}$  & $7.2^{0.3}_{0.3}$ & $1.5$ & $27.76^{0.03}_{0.01}$      \\
\\
MJD 54715 & $1.0$ & $0.76^{0.03}_{0.03}$  & $25.2^{3.6}_{0.3}$ & $2.139^{0.004}_{0.004}$ & $3.77^{0.04}_{0.04}$ & $10.02^{0.02}_{0.02}$  & $6.7^{0.3}_{0.3}$ & $1.5$ & $27.78^{0.03}_{0.02}$      \\
\hline
\hline
\end{tabular}
\vskip 0.4 true cm
\caption{Input model parameters corresponding to Figure~\ref{fits}. For each state we report: minimum, break and maximum
Lorentz factors of the electron distribution, low and high energy slope, magnetic field, electron density, radius of
emitting region and its Doppler factor.}
\label{param}
\end{table*}



This same SED of PKS 2155-304 had already been modelled in the original paper of~\cite{aha09}. On that occasion
the authors adopted a single zone SSC model, which they nevertheless observed to fail in producing a good fit to 
the SED and at the same time explain the daily observed variability. The lack of any counterpart to the X-ray 
variability, and the absence of signatures of any sort of change in 
the IC region of the spectrum made it difficult for them to advance a convincing model. Although a reasonable fit to the 
time-averaged SED had been achieved by postulating a particle energy distribution with two spectral breaks, the
final model could not account for the flux changes seen in X-rays. The latter had forced the authors to admit the necessity
of a two zone model, for which no physical constraints were available though.

In our analysis, the physical constraints for the two zone model were naturally and independently derived from the 
polarimetric analysis. Moreover, an exact counterpart for the X-ray variability was found in the active component
derived from the polarisation model of the source. This signature of an optical counterpart to the X-ray variability was 
not visible before because, corresponding to only 10\% of the total photometric flux, it was completely hidden within the 
rest of the jet. Therefore, this work serves also to warn that a number of variability events that oddly appear to  
have no counterparts in lower energies might have a hidden optical component whose small but variable flux is covered 
under the brighter emission of the extended jet.

We do not wish to go further into the physical interpretation of the results of this fit, but, to conclude, the reader should notice that the bulk of the broadband X-to-gamma-ray emission is produced by the variable component, singled out from the larger, steady jet component in the optical band by means of the polarimetric analysis. It was in fact this analysis which permitted us to disentangle the relation between the high and low-energy components of the SED, thus allowing for a coherent fit to the source's broadband spectral distribution to be achieved, emphasising the potential of the technique.

\section{Conclusions}

In this paper we have presented a new approach to the modelling of blazar SEDs which uses an analysis of the optical 
polarisation state of the source to provide additional, independent observational constraints to the SED parameters. In 
particular, the polarisation information can be used to motivate the necessity of a multi-zone model of the SED, being also
able to derive some of the fundamental properties of the two components, such as their relative flux and polarisation. 
Such information is then used to parameterise the inhomogeneous SED model, better constraining it. The technique was
illustrated by applying it to data from a campaign on PKS 2155-304, for which the complex variability behaviour and the 
lack of physical constraints for a two-zone model rendered it difficult explaining the SED. Based on the parameters derived
from the polarisation analysis we were able to derive a two-component model for the source which described its 
behaviour and temporal correlations in good detail, and a simple physical analysis of the results provides a self-consistent 
picture of the source.  

The fact that the polarisation analysis can disentangle the behaviour of a sub-component of the jet whose flux might only 
be a small fraction of the total emission is the key strength of the technique. In the particular case presented here,
the optical counterpart to the X-ray variability, previously interpreted as uncorrelated, was revealed to come from such
a zone responsible for not more than 10\% of the total optical output. This suggests that such kind of analysis 
might bear relevance to understanding MWL correlations and orphan flares in blazars. Of particular relevance to the 
general understanding of a blazar SED is the fact that our result suggests that during states of lower activity, a 
two-component model seems to give a better description of the emission. It is still to be verified what the application of
the technique could say about high-states, for which the strong MWL correlations hint toward the correlated signature of a 
single-zone. This question should be answered once we extend the analysis to a larger number of campaigns.

\section*{Acknowledgments}

Ulisses Barres de Almeida thanks the Brera Observatory, Merate, for their hospitality during the writing of this work. FT acknowledges financial contribution from grant PRIN-INAF-2011.


\label{lastpage}


\begin{thebibliography}{99}

\bibitem[\protect\citeauthoryear{Abdo et al.}{2010}]{abd10} Abdo et al. (LAT Collaboration) Nature 2010, 463, 919.
\bibitem[\protect\citeauthoryear{Abramowski et al.}{2012}]{abr12} Abramowski A. et al. (HESS Collaboration), 2012, A\&A, 539, 149.
\bibitem[\protect\citeauthoryear{Abramowski et al.}{2010}]{abr10} Abramowski A. et al. (HESS Collaboration), 2010, A\&A, 520, 83. 
\bibitem[\protect\citeauthoryear{Agudo et al.}{2011}]{agu11} Agudo I., Marscher A.P., Jorstad S.G. et al., 2011, ApJL, 735, L10.
\bibitem[\protect\citeauthoryear{Aharonian et al.}{2009}]{aha09} Aharonian F. et al. (HESS Collaboration), 2009, ApJL, 696, L150.
\bibitem[\protect\citeauthoryear{Aleksi\'{c} et al.}{2012}]{ale12} Aleksi\'{c} J. et al. (MAGIC Collaboration), 2012, A\&A,
542, 100.
\bibitem[\protect\citeauthoryear{Angel \& Stockman}{1980}]{ang80} Angel J.P.R. and Stockman H.S., 1980, ARA\&A, 18, 321.
\bibitem[\protect\citeauthoryear{Barres de Almeida et al.}{2010}]{bar10} Barres de Almeida U., Ward M.J. , Dominici T.P. et al., 2010, MNRAS, 408, 1778.
\bibitem[\protect\citeauthoryear{Bloom \& Marscher}{1996}]{blo96} Bloom S.D. \& Marscher A.P. 1996, 461, 657.
\bibitem[\protect\citeauthoryear{Brindle et al.}{1986}]{bri86} Brindle C., Hough J.H., Bailey J.A. et al., 1986, MNRAS, 221, 739.
\bibitem[\protect\citeauthoryear{D'Arcangelo et al.}{2009}]{dar09} D'Arcangelo F.D., Marscher A.P., Jorstad S.G. et al., 2009, ApJ, 697, 985.
\bibitem[\protect\citeauthoryear{Dominguez et al.}{2010}]{dom10} Dom\'{i}nguez A., Primack J.R., Rosario D.J. et al. 2010, MNRAS, 410, 2556. 
\bibitem[\protect\citeauthoryear{Fossati et al.}{1998}]{fos98} Fossati G., Maraschi L., Celotti A. et al., 1998, MNRAS, 299, 433.
\bibitem[\protect\citeauthoryear{Ghisellini et al.}{1998}]{ghi98} Ghisellini G., Cellotti A., Fossati G. et al., 1998, MNRAS, 301, 451.
\bibitem[\protect\citeauthoryear{Ghisellini \& Tavecchio}{2008}]{ghi08} Ghisellini G. and Tavecchio F., 2008, MNRAS, 386, 28.
\bibitem[\protect\citeauthoryear{Giannios et al.}{2009}]{gia09} Giannios D., Uzdensky D.A. and Begelman M.C., 2009, MNRAS, 395, 29.
\bibitem[\protect\citeauthoryear{Hagen-Thorn et al.}{2008}]{hag08} Hagen-Thorn, V.A., Larionov, V.M., Jorstad S.G. et al., 2008, ApJ, 672, 40.
\bibitem[\protect\citeauthoryear{Hinton \& Hofmann}{2009}]{hin09} Hinton J. and Hofmann W., 2009, ARA\&A, 47, 523.
\bibitem[\protect\citeauthoryear{Hughes et al.}{1989}]{hug89} Hughes P.A., Aller H. and Aller M., 1989, ApJ, 341, 68.
\bibitem[\protect\citeauthoryear{Jones}{1988}]{jon88} Jones T.W., 1988, ApJ, 332, 678.
\bibitem[\protect\citeauthoryear{Laing}{1980}]{lai80} Laing R.A., 1980, MNRAS, 193, 439.
\bibitem[\protect\citeauthoryear{Lovelace et al.}{2002}]{lov02} Lovelace R., Li H., Koldoba A.V., et al. 2002, ApJ, 572, 445. 
\bibitem[\protect\citeauthoryear{Lyutikov et al.}{2005}]{lyu05} Lyutikov M., Pariev V. \& Gabuzda D., MNRAS, 360, 869.
\bibitem[\protect\citeauthoryear{Mankuzhiyil et al.}{2011}]{man11} Mankuzhiyil N., et al. 2011, ApJ, 733, 14.
\bibitem[\protect\citeauthoryear{Mankuzhiyil et al.}{2012}]{man12} Mankuzhiyil N., Ansoldi S., Persic M. et al. 2012, ApJ, 753, 154. 
\bibitem[\protect\citeauthoryear{Maraschi et al.}{1992}]{mar92} Maraschi L., Ghisellini G. and Celotti A. 1992, ApJ Lett., 397, 5.
\bibitem[\protect\citeauthoryear{Marscher et al.}{2008}]{mar08} Marscher A.P., Jorstad, S.V., D'Arcangelo F.D. et al. 2008, Nature, 452, 966.
\bibitem[\protect\citeauthoryear{Marscher et al.}{2013}]{mar13} Marscher A.P. 2013, ApJ, in press. E-print: arXiv:1311.7665
\bibitem[\protect\citeauthoryear{Moore et al.}{1982}]{moo82} Moore R.L., Angel J.R.P., Duerr R. et al. 1982, ApJ, 260, 415.
\bibitem[\protect\citeauthoryear{Nakamura et al.}{2001}]{nak01} Nakamura M., Uchida Y. \& Hirose S. 2001, New Astron., 6, 61.
\bibitem[\protect\citeauthoryear{Sikora et al.}{1994}]{sik94} Sikora M., Begelman M. \& Rees M.J., 1994, ApJ, 421, 153.
\bibitem[\protect\citeauthoryear{Sironi et al.}{2011}]{sir11} Sironi L. \& Spitkovsky A., 2011, ApJ, 741, 39.
\bibitem[\protect\citeauthoryear{Tavecchio et al.}{1998}]{tav98} Tavecchio F., Maraschi L. and Ghisellini G., 1998, ApJ, 509, 608.
\bibitem[\protect\citeauthoryear{Tavecchio et al.}{2010}]{tav10} Tavecchio F., Ghisellini, G., Ghirlanda, Foschini, L. and  Maraschi L., 2010, MNRAS, 401, 1570.  
\bibitem[\protect\citeauthoryear{Urry \& Padovani}{1995}]{urr95} Urry P. and Padovani P., 1995, PASP, 31, 473.

\end{thebibliography}
\end{document}